\documentstyle[12pt]{article}
\begin{document}
\begin{flushright}
CU-TP-863
\end{flushright}
\begin{flushleft}
\vskip 20pt
{\Large\bf General Issues in Small-x and Diffractive
Physics\footnote{Work supported in part by the US Department of Energy
under Grant No. DE-FG02-94ER40819}$^,$\footnote{Talk given at ``Interplay
between Soft and Hard Interactions Deep Inelastic Scattering, Heidelberg,
Sept.29-Oct.,1997}}

\vskip 20pt
\noindent{A.H. Mueller}
\vskip 10pt
\noindent{Department of Physics, Columbia University, New York, N.Y. 10027}
\end{flushleft}
\vskip 40pt

{\bf Abstract}\ \ A selected review of topics at the border of hard and
soft physics is given.  Particular emphasis is placed on diffraction
dissociation at Fermilab and HERA.  Recently, significant differences between
diffraction dissociation at HERA and at Fermilab have become apparent.  This may
suggest that one already is reaching nonlinear (unitarity) effects which are
extending from the soft physics region into the semihard regime of QCD.

\section{Introduction}

 The focus in this paper is on the regime of hardness near the borderline
between hard and soft high energy collisions with a special emphasis on
searching for nonlinear QCD effects.  This is an opportune time for such a
discussion as there is now a significant body of complementary data from deep
inelastic scattering and from hadron-hadron collisions.  Perhaps the major object
of this review is to compare and contrast hadronic and deep inelastic
collisions, especially diffraction dissociation where there are major
differences between the hadronic and virtual photon initiated processes.

 Sec.2 is devoted to a brief review of some soft physics results on total
cross sections and diffraction dissociation.  Despite the fact that total cross
sections grow[1]  as
$(s/s_0)^\epsilon,$ with
$\epsilon
\approx 0.1,$ through the highest Fermilab energies it is argued that there
already is strong evidence of unitarity corrections being important from the
ISR to Fermilab energy regimes.  In particular, we suggest the lack of growth of
the single diffraction dissociation cross
section[2-5] as
due to the blackness of central proton-proton and proton-antiproton collisions.

 In contrast to a very weak growth of the single diffraction dissociation
cross section in hadronic collisions the energy dependence of virtual photon
diffraction dissociation appears to be significantly
stronge[6-7] than that expected from soft physics.  In Sec.3, we
argue that this may be due to blackness for the soft components of the virtual
photon's wavefunction and a subsequent dominance of the process by semihard
components as suggested  recently by Gotsman, Levin and
Maor[8]. If this is indeed
the case it means that for the first time one has evidence of unitarity
(nonlinear) effects extending into the semihard regime of QCD.

In Sec.4, we remark that the new D\O data[9,10] on rapidity gaps
between jets showing that the gap fraction of events decreases with energy may
be due to the same physics which slows the growth of the single diffraction
cross section.  If this decrease is indeed due to an energy dependent
(decreasing) survival probability a similar behavior would be expected for
comparable photoproduction data involving resolved photons, but such a decrease
would not be seen in deep inelastic scattering.

In Sec.5, we review BFKL  searches performed at H1, ZEUS and
D\O[9,11-13].  Each analysis, using two-jet  inclusive measurements 
at Fermilab and a forward jet measurement at HERA, finds some evidence for BFKL
behavior through an energy dependence which seems stronger than expected from
leading and next-to-leading order perturbation theory.  However, definitive
results have not yet been achieved.

In Sec.6, progress in calculating the next-to-leading corrections in
BFKL evolution is reviewed[14]. We may be near a
rather complete understanding of these corrections.  A preliminary estimate
suggests a substantial reduction of the BFKL pomeron intercept.

Sec.7 is devoted to a brief discussion of some topics involving nuclear
reactions.  Parametrizations of diffraction dissociation at HERA have been
successfully used to describe nuclear shadowing in fixed target deep inelastic
lepton-nucleus scattering[15].
$J/\psi$ production in proton-nucleus and nucleus-nucleus scattering continues
to be an important subject for research.  New phenomenological success in
describing all data except for Pb-Pb collisions by a simple absorption
model[16,17],
along with the suggestion that the Pb-Pb data may be qualitatively
different[18] have made
it even more important to connect
$J/\psi$ production and scattering more firmly with QCD.

\section{Clearly soft}

\subsection{Total cross sections}

Donnachie and Landshoff[1]
have shown that all high energy total cross sections for hadron-hadron
collisions can be written in the form

$$\sigma_{tot}=\sigma_0(s/s_0)^\epsilon + {\rm subleading\  terms}\eqno(2.1)$$

\noindent where $\sigma_0$ depends on the particular hadrons initiating the
collision and the subleading terms go to zero roughly like $(s/s_0)^{-1/2}.$ 
$s_0$ is an arbitrary scale factor while $\epsilon$ appears to be
universal and of size

$$\epsilon \approx 0.1.\eqno(2.2)$$

\noindent $1 + \epsilon = \alpha_p$ is the intercept of the soft pomeron in
Regge-language.  HERA data shows that (2.1) is also true for real photon-proton
collisions.  Of course a growth in energy as fast as that indicated in
(2.1) cannot persist at arbitrarily high energies because of limitations
required by the Froissart bound which does not permit total hadronic cross
sections to rise faster than $\ell n^2 s/s_0$ at asymptotic energies.  The fact
that the behavior indicated in (2.1) persists up to the highest Fermilab energy
region might seem to indicate that unitarity constraints, which are responsible
for the Froissart bound, are not yet effective in the presently available
energy region.  However, as observed long
ago[19], this is not the case. 
If one writes proton-proton total, inelastic and elastic cross sections in terms
of the S-matrix at a given impact parameter of the collision,
$S(b),$ as

$$\sigma_{in}=\int d^2b[1-S^2(b)]\eqno(2.3)$$

$$\sigma_{e\ell}=\int d^2b[1-S(b)]^2\eqno(2.4)$$

$$\sigma_{tot}=2\int d^2b[1-S(b)],\eqno(2.5)$$

\noindent then $S(b)$ depends on \ b\ roughly as indicated in Fig.1.  (We take
S(b) to be real for simplicity.)  

\indent For small values of impact parameter $S(b)$ is near zero for
proton-proton collisions already in the ISR and Fermilab fixed target energy
regime.  For proton-antiproton collisions $S(b)$ is quite small for $b<1 fm$ in
the Fermilab energy regime.  $S(b)$ near zero is a signal that unitarity
corrections are large though they are not so easy to see in the total cross
section because the radius of interaction is   expanding and the growth of
$\sigma_{tot}$ is mainly coming from that expansion.  Monte Carlo
simulations[20] of the dipole
formulation[21,22] of the
Balitsky, Fadin, Kuraev,
Lipatov[23,24] equation for the
academic case of heavy onium-heavy onium scattering show a similar phenomenon. 
For rapidities less than about 15 the BFKL equation is reliable for the total
onium-onium cross section, however, for small impact parameter collisions
important unitarity corrections are visible for rapidities of 6 units.

\subsection{Diffraction dissociation in hadron-hadron scattering}

\indent Single diffraction dissociation, illustrated in Fig.2, is given by 

$$x_P{d\sigma_{SD}\over dx_Pdt}\ =\ x^{2(1-\alpha_P(t))}f(M_x^2)\eqno(2.6)$$

\noindent is terms of soft pomeron exchange.  Integrating (2.6) over \ t\ and
over $x_P\leq 0.05,$ but excluding the proton state $M_x = M_p,$ one gets a
single diffractive cross section $\sigma_{SD}.$  From the Regge formalism one
expects $\sigma_{SD}$ to grow with\ $s$ \ as $s^{2\epsilon},$ but this is not
seen in the data as illustrated in Fig.3, which is a simplified version of the
more complete plot in Ref.5 where detailed  data points are shown.  A factor of
2 is included in Fig.3 to account for single diffractive dissociation of either
of the colliding protons (antiproton).  At Fermilab collider energies there
is a discrepancy of an order of magnitude between the Regge fit and the data.

\indent It seems clear that this discrepancy and the slow growth of $\sigma_{SD}$
with energy signal a breakdown of the Regge analysis when ${\sqrt{s}}\geq 20
GeV.$  I think this breakdown can be expressed in various equivalent ways.  (i) 
A low energies inelastic collisions induce, through unitarity, both elastic
scattering and diffractive dissociation.  However, as $S(b)$ goes to zero for
small and moderate  $b$ at high energies, these black regions of impact
parameter space only induce elastic scattering and not diffraction
dissociation.  Thus as one increases energy the elastic cross section  grows
rapidly while the diffraction dissociation cross section, coming from those
regions in impact parameter space where $S(b)$ is neither too close to zero or
too close to one, grows very slowly.  (ii)  In the Regge language one must
include multiple pomeron exchange in addition to the single pomeron exchange
which is valid at lower energy.  This multiple pomeron exchange gives absorptive
(virtual) corrections which slow the growth coming from single pomeron
exchange.  (iii) The ``gap survival'' probability[25,26] decreases with energy
compensating the growth due to single pomeron exchange.  Although gap survival
probability is a concept usually used for hard collisions I think the same idea
applies to single diffraction dissociation, at least in a heuristic way, in
hadron-hadron collisions.  It is likely that (i), (ii) and (iii) are just
different ways of saying the same thing.

\section{Deep inelastic lepton-proton scattering\newline 
analogs of the soft physics results}

\subsection{An ``elastic'' scattering amplitude}

\indent Recently, there has been an interesting suggestion as to how to test
unitarity limits in deep inelastic
scattering[27].  Of course there is
no Froissart bound for virtual photon-proton scattering, nevertheless, we have
become used to viewing the small-x structure function in terms of a high energy
quark-antiquark pair (possibly accompanied by gluons) impinging on the target
proton.  Although the quark-antiquark pair is not on-shell the time evolution of
the pair as it passes through the nucleon should be constrained by unitarity in
much the same way that a quark-antiquark pair coming from, say, a pion state
would be.

\indent To be more specific, view deep inelastic scattering in the rest system of
the proton and in the aligned jet (naive parton)
model[28,29].  At small x the virtual photon,
$\gamma^*(q),$ breaks up into a quark and antiquark pair long before reaching
the proton.  The relative transverse momentum of the quark and antiquark is
small, of hadronic size
$\mu\approx  350 MeV,$ while the longitudinal momenta are $q_z$ and $q_z\cdot
{\mu^2\over Q^2}$ respectively.  Because the relative transverse momentum is
small the  transverse coordinate separation of the quark and antiquark can be
expected to be on the order of a  fermi, and the resulting cross section with the
proton should be of hadronic size.  The smallness of the overall cross section
comes from the small probability, of order $\mu^2/Q^2,$ to find such an
aligned jet configuration in the wavefunction of $\gamma^*.$ (More probable
configurations in the $\gamma^*$ have smaller interaction probabilities. While
the aligned jet model cannot be expected to be a precise model of deeply
inelastic scattering it should reasonably characterize a significant portion
of deep inelastic events.)

\indent The inelastic reaction of this, longitudinally asymmetric,
quark-antiquark pair with the proton should produce a shadow quark-antiquark
pair in the final state.  If the center of the proton is relatively black to the
incoming quark-antiquark pair the shadow may be rather strong, as in the
hadronic case discussed above, and unitarity limits may aleady be reached at
present energies.  The outgoing quark-antiquark pair should show up  as a
diffractively produced state, of mass $M_x \approx Q,$ following the direction
of the $\gamma^*.$  Assuming that the scattering amplitude of the quark-antiquark
pair with the proton is imaginary one may  reconstruct this amplitude,
dropping an i, as

$$F(x,\b{b}) = \int d^2p e^{i\b{p}\cdot \b{b}}{\sqrt{{d\sigma_{SD}\over
d^2p}}}\eqno(3.1)$$

\noindent with $\b{b}$ the impact parameter of the collision and $\b{p}$ the
momentum transfer to the recoil proton.  ${d\sigma_{SD}\over d^2p}$ is the
single diffractive cross section for $M_x\approx Q.$

\indent In the present circumstance we do not have good control of the magnitude
of \ F\ near $\b{b}=0.$  However, if the proton is black for central collisions
one can expect $F(x,\b{0})$ to show little x-dependence. (Here x plays the role
that $s$ does for the hadronic collisions discussed above.)  The authors of
Ref.27 suggest looking at the b-dependence of

$$\Delta_{eff} = {d \ell n F(x,b)\over d \ell n 1/x}.\eqno(3.2)$$

\noindent Unitarity constraints can be expected to show up as smaller values
of $\Delta_{eff}$ near $b=0.$  More quantitatively, unitarity limits at $b=0$
would give

$$\Delta_{eff}(b=0) < 2(\alpha_P-1).\eqno(3.3)$$

\noindent This is a clever idea, and it will be interesting to see what the
data give.

\subsection{Large mass ${\gamma}^*$ diffractive dissociation}

\indent The traditional picture of large mass diffraction dissociation at
small  values of x is shown in Fig.4, where Dokshitzer, Gribov, Lipatov,
Altarelli, Parisi (DGLAP)[30-32] evolution takes one from the hard scale  
Q  to the soft scale
$\mu$ where a soft diffractive scattering, represented by soft pomeron exchange,
occurs.  In Fig.4, one assumes the DGLAP ordering

$$\mu^2 \approx \b{k}^2 << \cdot\cdot\cdot << \b{k}_2^2 << \b{k}_1^2 <<
Q^2.\eqno(3.4)$$

\indent However, this is a subtle process and it is worthwhile looking carefully
at the argumentation that leads to the size of $\b{k}^2$ at the lower end of
the DGLAP
evolution[8,33-35].  It is convenient to view that evolution proceeding from the
hard scale
\ Q\  toward softer scales, a direction opposite to that which is usually taken. 
In Fig.5, we illustrate the process in two steps:    The  left-hand part of the
figure shows the virtual photon wavefunction in terms of its quark and gluon
components.  As in Fig.4,   k  is supposed to be the softest gluon and $\Delta
x_\perp = 2/k_\perp$ gives the transverse size of the $\gamma^*$ state.  The
right-hand part of the figure gives the diffractive scattering part of the
process proceeding by gluon exchange from the proton interacting with the octet
dipole consisting of the gluon\ $k$ \ and the remainder of the $\gamma^*$
state.  Schematically, one may write

$$d\sigma_{SD} = {\rm flux}\  dP_r(k_\perp)[1-S(\Delta x_\perp = 2/k_\perp,
\b{b}, Y=\ell n 1/x)]^2d^2bdx_P\eqno(3.5)$$

\noindent where

$$dP_r(k_\perp) = {dk_\perp^2\over Q^2}\eqno(3.6)$$

\noindent is the probability that the lowest transverse momentum gluon have
momentum $k_\perp .$  Eq.(3.6) shows that gluons with small $k_\perp$ have a
small probability in the $\gamma^*$ wavefunction, analogous to what we found
earlier for low momentum quarks in the aligned jet model.  $1-S$ represents the
amplitude for a gluon having $k_\perp,$ along with the remainder of the
$\gamma^*$ wavefunction, to scatter elastically on the proton.   $b$  is the
impact parameter of the overall collision while $Y=\ell n 1/x$ is the rapidity
of the softest gluon with respect to the proton.  In $lowest\ order,$ two-gluon
exchange,

$$1-S \propto {x_PG(x_P,k_\perp^2)\over k_\perp^2}\eqno(3.7)$$

\noindent when $k_\perp$  is large and where an integration has been performed
over impact parameter, $\b{b}.$  Using (3.6) and (3.7) in (3.5) one sees,
dimensionally, that $k_\perp^2$ cannot be large and this is the logic that has
led theorists to take $k_\perp \approx \mu$ and use soft pomeron exchange for
the scattering amplitude, $1-S.$

\indent However, if  $S$  is near zero for $k_\perp = \mu$ and  for $\b{b} =
0,$ and this is not unreasonable since the  $S$  matrix is near zero for
small impact parameter hadron-hadron collisions, then it is apparent from
(3.5) and (3.6) that values of $k_\perp$  significantly larger than  $\mu$
will be important.  Indeed, the values of $k_\perp$ that will dominate large
mass single diffractive production are those values where  $S$ is near, but not
too close to, one.  This is the case since the probability in the $\gamma^*$
wavefunction is located in large $k_\perp-$ values.  The situation here is
quite different than for hadron-hadron collisions.  In hadron-hadron
collisions the wavefunction of the incoming hadron is, except for a very
small part, in the soft physics region. If the $S$-matrix is near zero for
central collisions then the inelastic reaction will feed into elastic
scattering as a shadow. In deep inelastic scattering at small-x when $S$ 
becomes black there will certainly be a similar phenomenon which occurs, and
which has been described in Sec.3.1, but, in addition, blackness in the small
$k_\perp$ region will allow higher values of $k_\perp$ to become effective thus
making the process semihard.

\indent If central impact parameter collisions of $\gamma^*$-proton collisions
are indeed black for $k_\perp \approx \mu$ then we would expect the x-dependence
of the single diffractive cross section, $x_P{d\sigma\over d x_P},$ to vary more
strongly with  \ x\ than suggested by the soft pomeron.  If one writes

$$x_P{d\sigma\over dx_P} \propto x^{-n}\eqno(3.8)$$

\noindent then both ZEUS[6] and
H1[7] now suggest that
$n \approx 0.4$ rather than the $n=2(\alpha_P-1) \approx 0.2$ predicted by the
soft pomeron.  If the ZEUS and H1 measurements hold up, and  \ n\ really is near
0.4 in the small $\beta$ region, then I think it becomes clear that semihard
physics is dominating the physics of large mass diffraction dissociation.  In
that case it is interesting to reexamine the ``elastic'' scattering analyses we
described in Sec.3.1 to see if the proposed procedure to measure blackness is
also destroyed by the dominance of gluons and quarks at higher $k_\perp$-
values.  Finally, it should be pointed out that there are already rather
detailed calculations of the phenomenon, at least for $q\bar{q}$ and $q\bar{q}g$
components of the $\gamma^*$ wavefunction, which arrived at a value $n \approx
0.5,$ not too far from
experiment[8].   Have we, for
the first time, actually seen the long sought after evidence for nonlinearity
(unitarity limits) in the semihard region of deep inelastic scattering?

\indent Before leaving this section,  it may be useful to again contrast
hadron-hadron scattering with $\gamma^*$-proton scattering.  In the purely
hadronic case the energy dependence of the single diffraction cross section is
much weaker than that predicted by the soft pomeron. We have interpreted this as
due to blackness in central proton-proton collisions which enhances the elastic
cross section but suppresses diffractive excitation.  In $\gamma^*$-proton
scattering, on the other hand, the energy dependence (x-dependence) is much
stronger than that predicted by the soft pomeron.  We have interpreted that also
as due to blackness of the soft components of the $\gamma^*$ now leading to an
enhanced role for the harder components of the $\gamma^*$-wavefunction and a
resulting stronger energy dependence of the cross section.

\section{Rapidity gaps between jets at Fermilab and HERA}

\indent Suppose one measures two jets having comparable but opposite 
transverse momentum along with the requirement that there be a rapidity 
gap between two jets.  One might hope that this would be a good process to
measure the hard (BFKL) pomeron as illustated in Fig.6[36].  There are, 
however, at least two worries with using this process to measure the hard
pomeron.  (i) The pomeron contribution to the hard quark-antiquark scattering
is[37].

$${d^{\sigma}\over dt} = (\alpha C_F)^4\  {\pi^3\over 4 t^2} {exp[2(\alpha_P-1)
\Delta Y]\over [{7\over 2} \alpha N_c\zeta(3) \Delta Y]^3}\eqno(4.1)$$

\noindent with $\Delta Y$ the rapidity between the two jets.  Here $\alpha_P-1=
{4\alpha N_c\over \pi} \ell n 2$ is the BFKL pomeron intercept.  The presence
of the factor $(\Delta Y)^3$ in the denominator in (4.1) strongly reduces
the effective growth of the cross section with $\Delta Y$ making the emergence
of the hard pomeron more difficult at moderate values of $\Delta Y.$  (ii) 
Perhaps more serious yet is the fact that the cross section for producing two
jets with a gap between them depends on the absence of a soft interaction
between the spectator parts of the proton and antiproton, the so-called gap
survival probability[36].  This
lack of factorization makes it difficult to make a precise comparison between
theory and experiment.

There is new data from  D\O[9,10] and an interesting new analysis
comparing the 1800 GeV data with that at 630 GeV.  If
$f_{gap}$ is the fraction of all two-jet events (separated by a given rapidity)
with a gap between them then D\O finds that

$${f_{gap}(630)\over f_{gap}(1800)} = 2.6 \pm 0.6_{stat.}\eqno(4.2)$$

\noindent for $\Delta Y \geq 3.8$.  Thus the gap fraction $decreases$ with
increasing energy.  While this number cannot be directly compared to BFKL
dynamics because $\Delta Y$ has been taken to be the same at the two energies,
while a BFKL test should have $\Delta Y(1800)-\Delta Y(630) = \ell n {1800\over
630},$ it does suggest that the survival probability has a rather strong energy
dependence  making BFKL tests more difficult in rapidity gap events.  It will
be interesting to see whether models of the gap survival probability can easily
accomodate the energy dependence in Eq.(4.2)[38].

At Fermilab the gap fraction is typically 0.01 while at HERA more like 0.07. 
The gap survival probability is much larger at HERA as is natural for
a point-like $\gamma^*.$ It would be interesting to have a HERA analysis similar
to that of\  D\O\  to see if the energy dependence of the gap fraction is
weaker, closer to x-independent, than at Fermilab.   With respect to the
D\O data the energy dependence of the gap fraction may be reflecting
exactly the same phenomenon as observed in the energy dependence of the single
diffraction cross section discussed in Sec.2.2.  While the inclusive two-jet
cross section increases at higher energies, because of the growth in the parton
densities, the energy dependence of the gap cross section is likely to be much
weaker because of the increasing blackness of central proton-antiproton
collisions as already seen in the single diffractive cross section.

\section{BFKL searches}

The hard (BFKL) pomeron or, equivalently, BFKL evolution shows up
simply only in single transverse momentum hard scale processes.  Thus in
hadron-hadron collisions or in deep inelastic scattering where a soft scale,
the size of the hadron (proton), is present a special class of events must be
taken in order to isolate BFKL dynamics.  Since this is generally very
difficult to do experimentally it is perhaps useful to remind the reader why
BFKL dynamics is so interesting for QCD and why it is worth the considerable
effort necessary to uncover it.

There are at least two important reasons why hard single scale high energy
scattering is interesting.  (i)  It is a high energy scattering problem that
may be soluble, or nearly soluble.  (ii) BFKL evolution leads to high parton
densities and thus into a new domain of nonperturbative, but weak coupling, 
QCD.  As parton distributions evolve from a momentum fraction $x_1$ to a
smaller momentum fraction $x_2$, all at a fixed transverse momentum scale,
BFKL dynamics gives the rate of increase of those (mainly gluon) densities. 
This evolution is illustrated in Fig.7.  When gluon densities reach a density
such that on the order of $1/\alpha$ gluons overlap, perturbation theory
breaks down and one enters a new regime of strong field, $F_{\mu\nu} \sim 1/g,$
QCD.  While it is unlikely that one can reach such densities at Fermilab or
HERA at truly hard transverse momentum scales one should at least be able to
see the approach to these high densities through BFKL evolution.

Inclusive two-jet cross sections at Fermilab and forward single jet inclusive
cross sections at HERA can be used to measure the BFKL
intercept[39-43]. 
These processes are illustrated in Figs.8 and 9 respectively where
$k_1$ and
$k_2$ represent measured jets.  In proton-antiproton collisions one chooses
$k_{1\perp}, k_{k_2\perp} > M, $ a fixed hard scale, while in deep inelastic
scattering $k_{1\perp}$ is chosen to be on the order of Q, the photon
virtuality.  For the hadron-hadron case

$$\sigma_{2-jet}= f(x_1, x_2, M^2) {e^{(\alpha_P-1)\Delta Y}\over \sqrt{\Delta
Y}}\eqno(5.1)$$

\noindent while for deep inelastic scattering

$$\sigma_{jet} = f(x_1,Q^2) {e^{(\alpha_P-1)\ell n x_1/x}\over \sqrt{\ell n
\ x_1/x}}\eqno(5.2)$$

\noindent with $x_1$ and $x_2$ being the longitudinal momentum fractions of the 
measured jets.  $\alpha_P-1= {4\alpha N_c\over \pi} \ell n 2$ and the $f$'s in
(5.1) and (5.2) are known in terms of the quark and gluon distributions of the
proton and antiproton.  In (5.1) $\Delta Y$ is the rapidity difference between
the  two measured jets.  One can get a measurement of $\alpha_P-1$ in (5.1) by
varying $\Delta Y$ with $x_1, x_2$ and $M^2$ fixed, and this can be done at
Fermilab by  comparing the inclusive two-jet cross section at different
incident energies.  In (5.2) one can measure $\alpha_P-1$ by varying\ $x$\ for
fixed \ $x_1$ and $Q^2.$

Sometime ago H1[11,12] presented an analysis showing
$\sigma_{jet}$ increasing by about a factor of four as $x$  goes from about
$3x10^{-3}$ to about $7x10^{-4}$ for $k_{1\perp} > 3.5 GeV.$  This is a growth
quite a bit faster than given in conventional Monte Carlos and much faster than
the growth from single gluon exchange between the measured jet and the
quark-antiquark pair coming from the virtual photon.  The growth is comparable
to that given in (5.2), for $\alpha_P-1 \approx 1/2,$ however, the comparison is
not completely convincing because a comparison of partonic energy dependences,
from (5.2),  with hadron final states is not very reliable when $k_{1\perp}$ is
as small as in the $H1$ analyses.

Recently ZEUS[13] has completed an
analysis of this process.  Since the ARIADNE Monte Carlo gives a good fit to the
ZEUS data this Monte Carlo is used to unfold the hadronization and thus get a
better comparison with BFKL evolution.  The data agree much better with BFKL
evolution than with the Born term or with next-to-leading order QCD
calculations.  A definitive comparison with BFKL dynamics is hindered by the
lack of ability to include hadronization corrections along with the BFKL
evolution.  One can hope that the situation will soon improve in this regard.

A new D\O[9] comparing 1800 GeV and 630 GeV data for
$k_{1\perp}, k_{2\perp} \geq 20 GeV$ gives $\alpha_P=1.35 \pm 0.04 (stat) \pm
0.22$ (syst) when (5.1) is used to fit the data.  The strength of the D\O
analysis is that $k_\perp > 20 GeV$ which makes uncertainties due to jet
definition minimal.  Weaknesses of the analysis are the large systematic error
and the smallness of $\Delta Y,$ equal to 2, at the lower energy.  We can hope
that the systematic errors will come down in the near future.

Overall, I think the BFKL searches are encouraging but not yet definitive. The
fact that all three analyses suggest a strong increase with energy of reliable
quantities for isolating BFKL effects is certainly positive. An attempt will
also be made to measure $\alpha_P-1$ at LEP[44] in the next year by measuring the
$\gamma^*-\gamma^*$ total cross section.  This is a very clean process, although
the cross section is rather small.

\section{Higher order corrections to BFKL evolution}

In general in QCD next-to-leading corrections are very important.  It is only
after next-to-leading corrections have been calculated that scales have a real
meaning and normalizations can be trusted.  In the case of BFKL evolution the
next-to-leading corrections are also important to show that, in principle, 
corrections to the BFKL answer can be calculated, thus making single scale
high  energy scattering systematically calculable in QCD.

There has been a
long
program[14,45-47],
led by the work of V. Fadin and L. Lipatov, to calculate the next-to-leading
corrections to BFKL evolution and it now appears that  program may be coming to
completion.  When the work is finished one should get the next correction to
$\alpha_P$ as well as next-to-leading resummations for anomalous dimension and
coefficient functions.  If one writes

$$\alpha_P = {4 N_c\over \pi} \ell n 2 \alpha(Q)[1-c \alpha(Q)]\eqno(6.1)$$

\noindent then there is the suggestion\   $c$\   may be near 3, a very large
correction, although there is some work yet to be done before one can accept
this number with confidence[14].

For the anomalous dimension matrix one writes

$$\gamma_n = \sum_{i=1}^\infty \gamma_{n i}^{(0)}\left[{\alpha N_c\over
\pi(n-1)}\right]^i + \alpha \sum_{i=1}^\infty \gamma_{n i}^{(1)}\left[{\alpha
N_c\over \pi(n-1)}\right]^i + \cdot \cdot \cdot\eqno(6.2)$$

\noindent where the first series represents the leading order (BFKL) answer. 
We should soon know the second series, the constants $\gamma_{n i}^{(1)}$ along
with similar terms for the coefficient functions.  When the BFKL corrections
are known at next-to-leading order we should reap several benefits.  (i)  A
better understanding of the importance of BFKL (resummation) effects in $\nu
W_2$ should be possible.  Recall, that as a two-scale process BFKL dynamics
does not directly govern the small-x behavior of $\nu W_2.$  However, BFKL
effects are certainly present and can be systematically included through
resummations such as the one indicated in (6.2).  When the next-to-leading
corrections are known we should begin to get a reliable indication of the
importance of these resummation effects in the HERA regime.  (ii)  The
next-to-leading resummations should help us to better understand where the
operator expansion is valid in small-x physics, that is at what \ $x$ \ and
$Q^2$ are coefficient and anomalous dimension functions sufficiently safe from
diffusion effects to be reliably calculated
perturbatively[47,48].

\section{Nuclear reactions}
\subsection{Nuclear shadowing in deep inelastic scattering}

Nuclear shadowing in deep inelastic scattering is known to be a leading twist
phenomenon[29] and thus
dominated by soft physics, at least at current x-values.  In the DGLAP formalism
shadowing effects are put into initial parton distributions.  However, when
shadowing is not too strong it can be calculated from diffractive deep inelastic
scattering using the Gribov-Glauber formalism as illustrated in Fig.10.  In that
figure the left-hand part represents the imaginary part of the forward
$\gamma^*$-scattering amplitude which, by the optical theorem, is equal to the
$\gamma^*$-A total cross section or, equivalently, $\nu W_2.$  The first term
on the right-hand side of Fig.10 represents the incoherent scattering off the 
A  nucleons in the nucleus, the nucleons being labeled by $N_i.$  The second
term on the right-hand side of the figure represents the double scattering term
which is dominated by diffractive scattering, off nucleon $N_i$ in the
amplitude and $N_j$ in the complex conjugate amplitude.  So long as shadowing
correction are not too large it should not be necessary to go beyond the
double scattering term.  Indeed, the double scattering term can be obtained from
diffractive data at HERA while triple and higher scattering terms would involve
the scattering of partonic systems in the nucleus, terms which cannot be reliably
determined.  Using a parametrization which fits the HERA diffractive data, a
pretty good description of fixed target deep inelastic scattering of nuclei is
obtained with the double scattering term giving a shadowing correction of about
the right size[15].

\subsection{$J/\psi$ production in proton-nucleus and nucleus-nucleus
collisions}

Recently there has been much interest, and excitement, about the NA50
data[18]. 
In a nutshell one can summarize the  experimental situation as follows: (i) All
P-A and A-A collisions, except for Pb-Pb,
 {\em look\  like}  $J/\psi$ production in proton-proton collisions with
absorptive final state interactions corresponding to a $``J/\psi''$ cross
section with nucleons of
7mb[16,17].  (ii)  Pb-Pb central
collisions have a
$J/\psi$ cross section which is significantly suppressed with respect to
(i)[18] .

Kharzeev and Satz[49] have 
suggested a
picture that considers the system moving though the nuclei, after the hard
collision which produces the
$c\bar{c}$\  pair, to be a \  $(c\bar{c}g)$ color singlet system which, if it
suffers no reaction with the nuclear medium, turns into a $J/\psi$ after the
$(c\bar{c}g)$ system has passed through the material.  The $(c\bar{c}g)$ system
is supposed to have a size comparable to that of the $J/\psi,$ but, because
of the fact that it looks like an octet dipole formed from $(c\bar{c})_8,$ and a
gluon a cross section  of 7mb is natural.  This is an interesting picture,
however, there are a lot of unanswered questions.  (i) Where does the gluon in
the $(c\bar{c}g)$ system come from?  Does it come from the hard scattering or is
it part of the gluon distribution of the incident hadron or nucleus?  If it is
the latter does this
$enhance\ J/\psi$ production in nuclear collisions because there are so many
more spectator gluons at the impact parameter of the collision.  (ii)  How does
the gluon know what size system to form with the $c\bar{c}$ since the $(c\bar{c}g)$
does not interact, due to a slowing down of the rate of interaction for high
velocity states, while traversing the material?  Why should the relevant size
for the $\psi$ and $\psi^\prime$ be the same?  While a very interesting, and
successful, phenomenology has developed around this picture it is important to
determine whether the whole picture is reasonable from a QCD point of view.

\newpage
\noindent{\Large\bf REFERENCES}

\vskip 10pt
\begin{enumerate}

\item A. Donnachie and P.V. Landshoff, Phys.
Lett. B296 (1992)
  				 227, and references therein.

\item	D. Bernard et al., UA4 Collaboration, Phys. Lett. B186(1987)
227.

\item						F. Abe et al., CDF Collaboration, Phys. Rev.D50
(1994) 5535.

\item					N.A. Amos et al., E710 Collaboration, Phys. Lett.
B301 (1993) 313.

\item						K. Goulianos, Phys. Lett. B358 (1995) 379.

\item						ZEUS Collaboration contribution to EPS 1997.

\item						H1 Collaboration contribution to EPS 1997.

\item					E. Gotsman, E. Levin and U. Maor,
Nucl. Phys.B493 (1997) 354.

\item			  A. Goussiou in presentation for the
D\O Collaboration at EPS {1997}. 

\item				A. Brandt in presentation for the D\O
Collaboration at
 														``Interplay between Soft and Hard
Interactions in Deep
 Inelastic Scattering,'' Heidelberg, Sept.29-31,
1997.

\item			 S. Aid et al., H1 Collaboration, Phys. Lett. B356 (1995)
118.

\item		 M. Wobisch for the H1 Collaboration in DIS1997.

\item			 ZEUS Collaboration contribution to EPS-see ref.6
1997.

\item			M. Ciafaloni, hep-ph/9709390.

\item			 A. Capella, A. Kaidalov, C. Merino, D. Pertermann, and
J.
 														Tran Thanh Van, hep-ph 9707466; A. Bialos,
W. Czyz and 
 W. Florkowski, TPJU-25/96.

\item			C. Gerschel and J. H\"{u}fner, Phys.Lett. B207
(1988); Zeit. Phys.
  													C56 (1992) 391.
						
\item			 D. Kharzeev, Nucl. Phys. A610 (1996) 418c.

\item			 M. C. Abrew et al., NA50 Collaboration, Nucl. Phys.
A610
  				(1996) 404c.

\item			 U.Amaldi and K.R. Schubert, Nucl. Phys.B166(1980)
301.

\item			 G.P. Salam, Nucl. Phys. B461  (1996) 512.

\item			 A.H. Mueller, Nucl. Phys. B415 (1994) 373; A.H.
Mueller and
 														B. Patel, Nucl. Physics B425 (1994)
471.

\item		 N.N. Nikolaev, B.G. Zakharov and V.R. Zoller, JETP
Lett.	
 														59 (1994) 6.

\item			Ya.Ya. Balitsky and L.N. Lipatov, Sov.J.
Nucl.Phys.28 (1978) 22.

\item			E.A. Kuraev, L.N. Lipatov, and V.S. Fadin,
Sov.Phys. 
 													JETP 45 (1977) 199.

\item			J.D. Bjorken, Phys. Rev.D47 (1993) 101.

\item			 E. Gotsman, E.M. Levin and U. Maor,
Phys. Lett. B309
 				(1993) 199.

\item			 B. Kopeliovich, B. Povh and E. Predazzi,
hep-ph/9704372.

\item			J.D. Bjorken in Proceedings of the
International Symposium Electron and Photon Interactions at High Energies,
Cornell{(1971)}.

\item				 L.L. Frankfurt and M. Strikman,
Phys.Rep.160(1988) 235.

\item				Yu. L. Dokshitzer, Sov. Phys. JETP, 73
(1977) 1216.

\item		V.N. Gribov and L.N. Lipatov, Sov.J. Nucl.
Phys. 15
 (1972) {1978}.
 														
\item		G. Altarelli and G. Parisi, Nucl. Phys. B126
(1977) 298.

\item			 A. Hebecker, hep-ph/9702373.

\item		 M. W\"{u}sthoff, hep-ph/9702201.

\item			J. Bartels and M. W\"{u}sthoff in DIS 97.

\item			 J.D. Bjorken, Int. J. Mod.
Physics A7(1992) 4189.

\item				A.H. Mueller and W.-K. Tang, Phys.
Lett. B284 (1992) 123.

\item			O.J. Eboli, E.M. Gregores and F. Halzen,
hep-ph/9708283.

\item 			A.H. Mueller and H. Navelet, Nucl.
Phys. B282 (1982) 727.

\item			 W.-K. Tang, Phys. Lett. B278 (1991) 363.

\item		 J. Bartels, A. De Roeck and M. Loewe, Z.
Phys. C54 (1992) 635.

\item			 J. Kwiecinski, A.D. Martin and P.J.
Sutton, Phys. Rev.D46
 														(1992) 921.

\item	   V. Del Duca, hep-ph/9707348.

\item				A. De Roeck, private
communication.

\item			 V.S. Fadin and L.N. Lipatov,
Yad.Fiz.50 (1989) 1141;
 														Nucl.Phys. B406 (1993) 259;
Nucl.Phys.B477 (1996) 767.

\item		 V.S. Fadin, R. Fiore and M.I. Kotsky,
Phys. Lett.
 														B359 (1995) 181, B387 (1996) 593,
B389 (1996) 737.

\item		 G. Camici and M. Ciafaloni, Phys. Lett.
B386  (1996) 341, B395
											(1997)	 118; Nucl. Phys. B496 (1997) 305.

\item		 A.H. Mueller, Phys. Lett. B396
(1997) 251.

\item				D. Kharzeev and H. Satz, Phys. Lett. B366
(1996) 316.

\end{enumerate}

\end{document}